# Fundamental Building Blocks for The Design of A Single-electron Nanoelectronic Processor

Georgios Zardalidis[*] and Ioannis Karafyllidis
Democritus University of Thrace
Department of Electrical and Computer Engineering
671 00 Xanthi, GREECE
*Corresponding author, e-mail: gzardali@ee.duth.gr

**Abstract** - A single-electron random access memory array (RAM) and a single-electron universal Fredkin gate are designed and simulated. The universality of the Fredkin gate in combination with the RAM gives the potential of the realization of an elementary single-electron nanoelectronic processor.

## I. INTRODUCTION

Single-electronics is one of the emerging nanoelectronic technologies that deal with the control of transport and position of a single or a small number of electrons. The fundamental physical principle of single-electronics is the tunnelling effect and the phenomenon of Coulomb blockade.

Several single-electron circuits have been recently proposed in the literature: single-electron memories [1], inverters and pumps [2], majority gates [3], logic gates [4], [5], half adders [6] and adders [7]. The need for computer-aided design and simulation of single-electron circuits has long been recognized [8]. Several simulators have been implemented to support single-electron circuit design [9], [10]. Nevertheless the development and fabrication of single electron devices has already taken place [11- [13]. More about single electron circuits can be found in [14], [15].

One of the promising single-electron memory cells is the electron trap [16]. This cell presents a high energy barrier to electron transport because of thermal fluctuations, is not sensitive to random background charge, and has been recently fabricated [17].

The most difficult hurdle for constructing single-electron memory arrays is the selective *read* and *write* operations. One of the aims of this work is to tackle this problem.

The Fredkin gate, known also as Fredkin-Toffoli gate, has been proposed by Fredkin and Toffoli in 1982 [18] and performs conditional crossover of two data bits according to the values of a control bit. The Fredkin gate is a computationally universal gate, i.e. any Boolean function can be implemented using only by Fredkin gates.

A Single-electron Fredkin gate (SEF-gate) would be very useful, because digital single-electron circuits that use only one type of gate would be very much easier to fabricate than single-electron circuits that use various types of gates.

## II. OPERATION OF THE BASIC SINGE-ELECTRON MEMORY CELL

Figure 1(a) shows the circuit of the basic single-electron memory cell that will be used throughout this work, i.e. the electron trap. The electron trap circuit comprises six islands $N$, bounded by one capacitor, $C_1$, and six tunnel junctions $J$. The tunnel junctions are identical. When a positive voltage $V_g$ is applied, an electron or a small number of electrons is transported from the ground to island $N_1$. The presence of an excess electron or a small number of excess electrons at $N_1$ corresponds to logical "1" and the absence to logical "0". Figure 1(b) shows the symbol of the electron trap. Here junctions $J_1$ - $J_6$ and islands $N_2$ - $N_6$ are absorbed into the symbol *MTJ*, that stands for "multiple-tunnel junction".

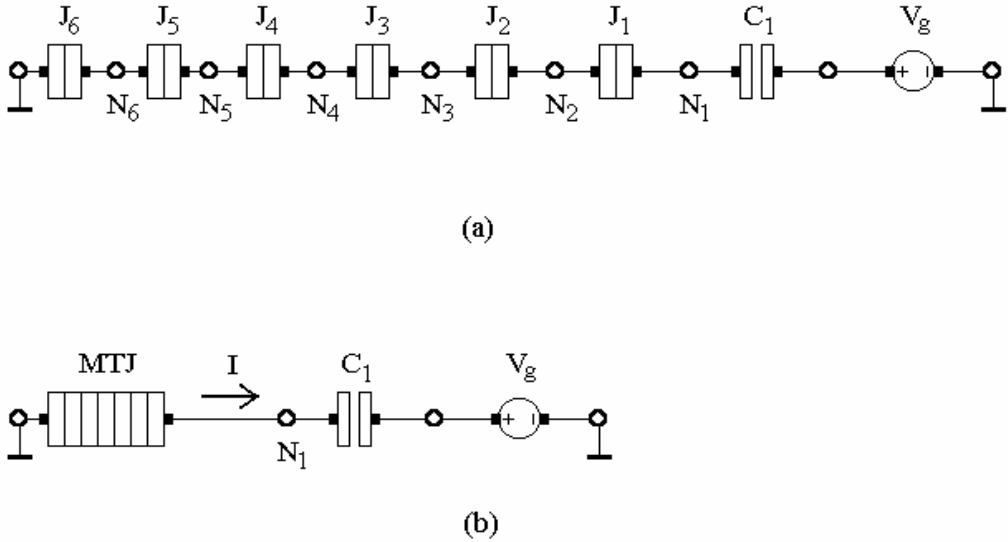

**Figure 1.** (a) The circuit of the electron trap, (b) the symbol of this circuit.

The simulated operation of the basic memory cell of Figure 1 can be described as follows. Initially, the voltage $V_g$ is zero and there are no excess electrons at the island $N_1$. After some time $V_g$ becomes positive and reaches the value of *0.6 V*. At this time an electron is transported from the ground to $N_1$ and causes the negative current pulse. Although $V_g$ remains at *0.6 V* for some time, no more electrons are transported to $N_1$, because of the Coulomb blockade. After that, the value of $V_g$ becomes zero again and remains zero for some time, but the electron remains in $N_1$, because it can not surmount the potential barrier imposed by junctions $J_1$ - $J_6$ and escape to ground. Thus the logical "1" has been written in the memory cell and this logical value is kept at the cell although the value of $V_g$ has returned to zero.

To write the logical value "0" in the cell, which currently keeps the logical value "1", $V_g$ becomes negative and reaches the value of *-0.4 V*. At this time the excess electron is transported from $N_1$ to the ground and causes the positive current pulse. After that the value of $V_g$ remains at *–0.4 V* for some time and then becomes zero again. Thus the logical "0" has been written in the memory cell.

From the simulation becomes clear that the electron trap has two stable states, corresponding to energy levels $E_A$, (logical "0") and $E_F$ (logical "1"). To change the state of the electron trap from "0" to "1" the electron has to surmount an energy barrier equal to ($E_D$ - $E_A$), whereas to change the state from "1" to "0" the electron has to surmount an energy barrier equal to ($E_D$ - $E_F$). Since ($E_D$ - $E_A$) > ($E_D$ - $E_F$) a smaller absolute value of $V_g$ is required (i.e. *|0.4| V*) in order to drive the circuit from state "1" to state "0" than the absolute value required for driving the circuit from state "0" to state "1" (i.e. *|0.6| V*).

To read the content of the memory cell an electrometer, which detects the presence or absence of excess electrons at the island $N_I$, is used. This electrometer has already been fabricated [17].

### III. THE SINGLE-ELECTRON RANDOM ACCESS MEMORY ARRAY

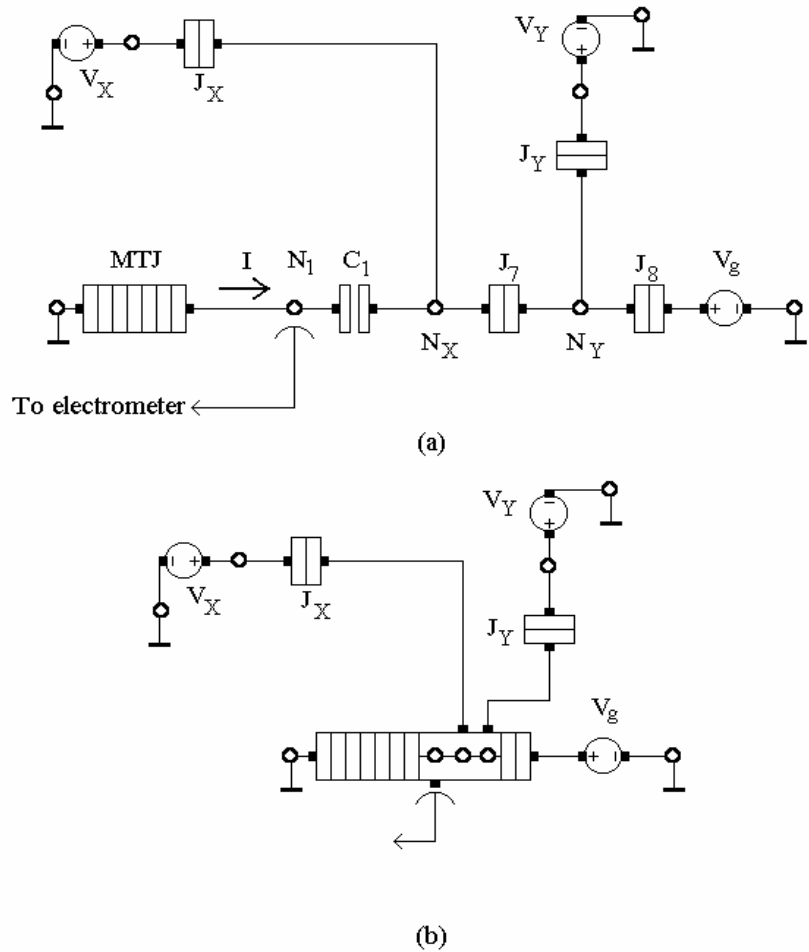

**Figure 2.** (a) Electron trap in which the access of $V_g$ to the island $N_I$ is controlled by two voltages, $V_X$ and $V_Y$, which are applied to nodes $N_X$ and $N_Y$ through junctions $J_X$ and $J_Y$, respectively, (b) the symbol of this circuit.

The most difficult hurdle for constructing single-electron memory arrays is the selective *read* and *write* operations. The aim of this work is the design and the simulation of the operation of a single-electron random access memory array using the electron trap as the basic memory cell. The basic cells will be organized as a *xy* matrix and two voltages, $V_X$ and $V_Y$, will be used to control the horizontal (*x*-direction) row cells and the vertical (*y*-direction) column cells, respectively.

*A. Selective writing*

Figure 2(a) shows an electron trap in which the access of $V_g$ to the island $N_1$ is controlled by two voltages, $V_X$ and $V_Y$, which are applied to nodes $N_X$ and $N_Y$ through junctions $J_X$ and $J_Y$, respectively. The tunnel junctions $J_7, J_8, J_X,$ and $J_Y$ are identical. The resistance of each one of these junctions is $10^5$ *Ohm* and the capacitance $10^{-18}$ *F*. Figure 2(b) shows the symbol of the electron trap, which comprises junctions $J_7$ and $J_8$, and islands $N_X$ and $N_Y$.

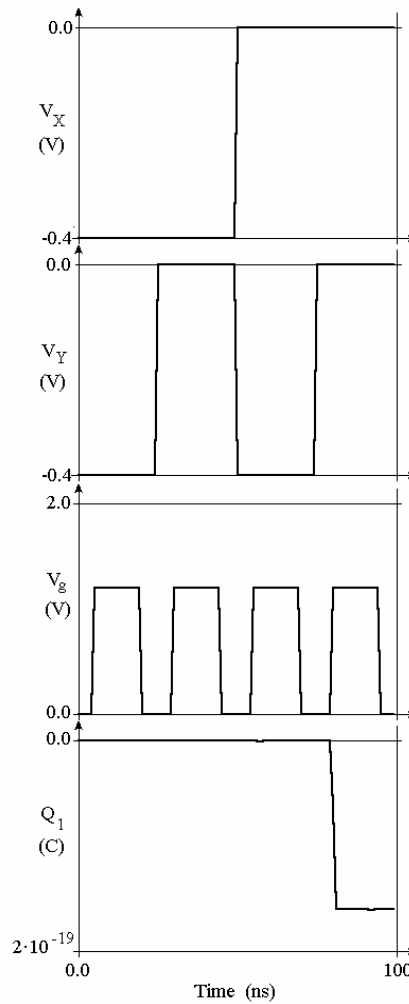

**Figure 3.** Simulation results for selective writing of logical "1" to the memory cell.

$V_X$ and $V_Y$ can take only two values, *0.0 V* and *–0.4 V*. $V_g$ can access the island $N_1$ to write the logical "0" or "1" only if $V_X = V_Y = 0.0\ V$. In any other case (i.e. $V_X = V_Y = -0.4V$, or $V_X = -0.4V$ and $V_Y = 0.0\ V$, or $V_X = 0.0V$ and $V_Y = -0.4\ V$) the memory cell can not be accessed. If any of the two selection voltages is equal to *–0.4 V*, the circuit free energy increases when electrons tunnel in or out of the island and thus, this tunneling event is suppressed. This is shown in Figure 3 where the simulation results of the memory cell access are presented.

Initially the memory cell stores the logical "0", i.e. no excess electrons are present at island $N_1$. As shown in Figure 3, all possible combinations of the $V_X$ and $V_Y$ values are applied to islands $N_X$ and $N_Y$. An attempt to write the logical "1" to the memory cell takes place at each one of the four possible combinations of $V_X$ and $V_Y$ values, by raising the value of $V_g$ to *1.2 V*. This value is larger than the *0.6 V* of the case of Figure 1, because $V_g$ now drives a larger capacitance. The simulation results of Figure 3 show that this attempt is successful only in the case of $V_X = V_Y = 0.0\ V$.

*B. The single-electron memory array*

A single-electron memory array in which selective writing and reading is possible can be constructed using as basic cell the cell presented in Figure 4. The selective writing operation has been shown in the last section, whereas, the reading operation is performed by sensing the charge at memory islands using electrometers.

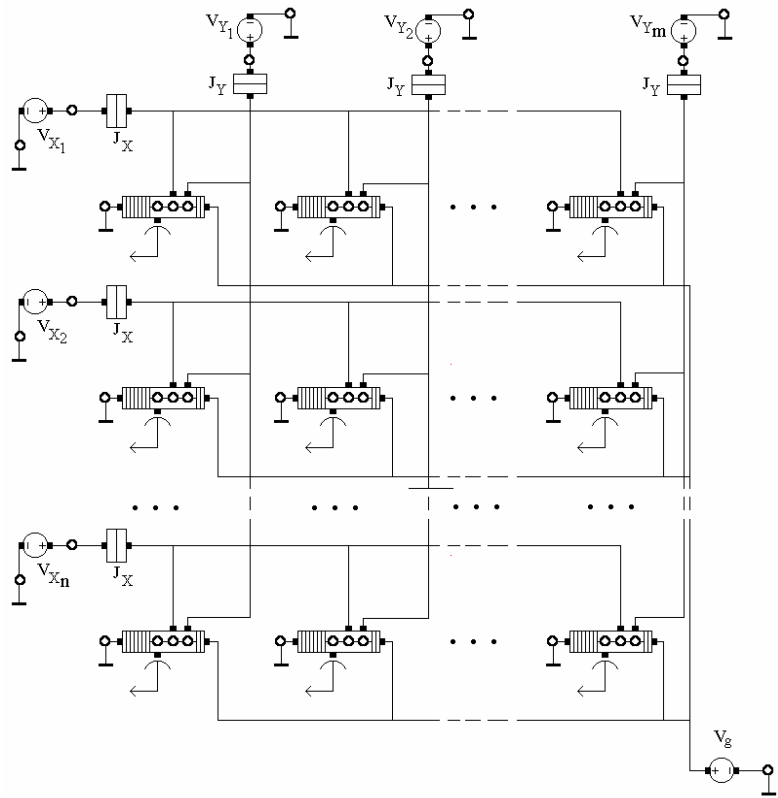

**Figure 4.** An (*n* x *m*) single-electron memory array in which the symbol of Figure 2(b) has been used.

Figure 4 shows an (*n* x *m*) single-electron memory array in which the symbol of Figure 4(b) has been used. Voltage lines $V_{Xi}$ and $V_{Yj}$ with *i = 1, 2, 3, …n* and *j = 1, 2, 3,*

...*m*, control the access to the memory cells. To access the (*k*, *l*) cell, all voltages $V_{Xi}$, and $V_{Yj}$ are set to –*0.4 V* except $V_{Xk}$ and $V_{Yl}$ which are set to *0.0 V*.

## IV. THE SINGLE - ELECTRON FREDKIN GATE

*A. The single electron circuit*

A SEF-gate is a single-electron logic circuit that has the same logic operation as the classical F-gate but is constructed using single-electron circuitry. The SEF-gate presented here, is reversible in the sense that the input data can be deduced if the output data are known, but inputs cannot be physically regenerated by applying outputs as inputs, because of the non-reversibility of the electron tunneling process.

The circuit of the single-electron SEF-gate is shown in Figure 5. The circuit comprises sixteen islands *Na*, *Nb*, *Nc* and *N1* through *N13*. The islands *Na*, *Nb* and *Nc*, are the output islands. *Na* is the output coming from the controlling input and *Nb* and *Nc* are the exchange outputs. Twenty-two junctions *J1* through *J22* bound the islands. Junctions *J4* and *J13* are less transparent in order to prevent electron transport from the ground (*Vss*) to islands *Na* and *N8* and vise versa, while junctions *J8* and *J19* are less transparent in order to prevent electron transport and charge oscillation to islands *Nb* and *Nc*. The voltages *Vdd1*, *Vdd2*, *Vdd3* and *Vdd4* are constant at -0.1V, 0.5V, -0.1V, and -0.11V respectively.

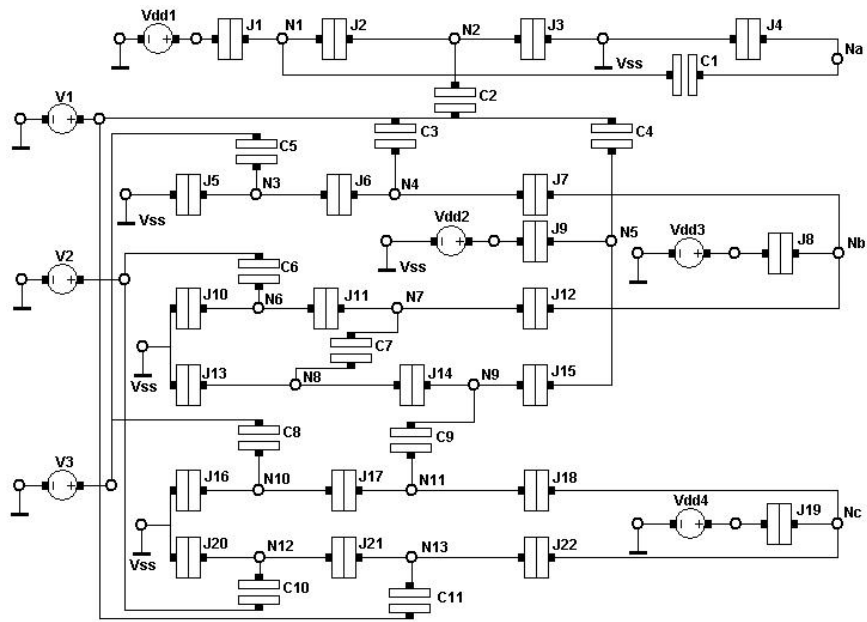

**Figure 5.** The circuit of the single-electron F-gate.

The voltage source *V1* is the controlling input of the SEF-gate, while the voltage sources *V2* and *V3* are the target inputs. These input voltages can take only two values 0.0Volt which corresponds to the logic "0", and 0.1Volt which corresponds to the logic "1". The input voltage *V1* is applied to islands *N2*, *N4*, *N5* and *N13* through capacitors *C2*, *C3*, *C4* and *C11* respectively. The input voltage *V2* is applied to islands *N6* and *N12* through capacitors *C6* and *C10* respectively. Finally, the input voltage *V3* is applied to islands *N3* and *N10* through capacitors *C5* and *C8* respectively. The capacitors *C1* through *C11* have a capacitance of $10^{-18}$F. The output signals of the F-gate are taken from islands *Na*, which corresponds to the controlling input *V1*, and *Nb* and *Nc*, which are the outputs that correspond to the exchanging inputs *V2* and *V3*, respectively. The presence of positive charge at the output islands corresponds to logic "1", whereas no charge corresponds to logic "0".

*B. Gate operation*

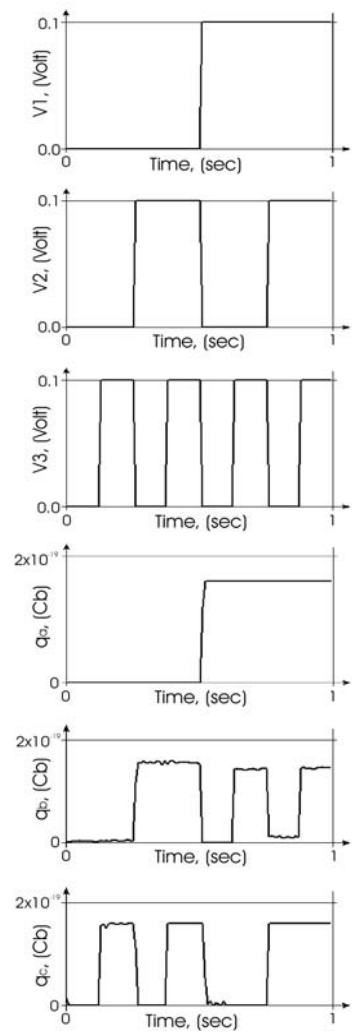

**Figure 6.** Operation of the SEF-gate. (a) time variation of controlling input voltage *V1*, (b) time variation of target input voltage *V2*, (c) Time variation of target input voltage *V3* (d) time variation of the charge at the controlling output island *Na*, (e)-(f) time variation of the charge at the exchange output islands *Nb*, *Nc* respectively.

The logic operation of the SEF-gate is shown in Figure 6. Figures 6(a), 6(b) and 6(c) show the time variation of the input voltages *V1, V2* and *V3*, respectively. The inputs are piece-wise constant and apply all possible combinations of logic "0" and "1" to the circuit. Figures 6(d) 6(e) and 6(f) show the time variation of the charge $q_a$, $q_b$ and $q_c$ at the output islands *Na*, *Nb* and *Nc* respectively. The results from the graphs of Figure 6, compose the truth table of the SEF-gate. The output transition from logic "0" to logic "1" and vice versa does not drive the circuit to instability. The presence of charge on the output islands can be detected and transferred to circuits connected to the SEF-gate output using a sense amplifier

## VI. CONCLUSIONS

The design of a single-electron random access memory array and a Single-electron Fredkin gate have been presented. The memory array utilizes a basic single-electron memory cell that has been recently fabricated. Simulation, using a Monte Carlo simulator, showed that selective writing is possible in this array, whereas, the reading operation can be performed by sensing the charge at memory islands using electrometers. The SEF-gate would be very useful, because digital single-electron circuits that use only one type of gate would be very much easier to fabricate than single-electron circuits that use various types of gates.

It's universal nature gives the SEF gate the ability and the advantage of being used as a fundamental building block for the implementation of a variety of logical circuits, which in combination with the SE RAM can lead to an elementary processor.